\documentclass[aps,pre,twocolumn,groupedaddress,superscriptaddress,showpacs]{revtex4-1}

\pdfoutput=1

\usepackage{graphicx}
\usepackage{color}
\usepackage{amsmath}
\usepackage{amsfonts}
\usepackage{amssymb}
\usepackage{dcolumn}

\begin{document}

\title{Corrections to Scaling for Watersheds, Optimal Path Cracks, and Bridge Lines}

\author{E.~Fehr}
%\email{ericfehr@ethz.ch}
\affiliation{IfB, ETH Z\"urich, CH-8093 Z\"urich, Switzerland}

\author{K.~J.~Schrenk}
%\email{jschrenk@ethz.ch}
\affiliation{IfB, ETH Z\"urich, CH-8093 Z\"urich, Switzerland}

\author{N.~A.~M.~Ara{\'u}jo}
\email{nuno@ethz.ch}
\affiliation{IfB, ETH Z\"urich, CH-8093 Z\"urich, Switzerland}

\author{D.~Kadau}
%\email{dkadau@ethz.ch}
\affiliation{IfB, ETH Z\"urich, CH-8093 Z\"urich, Switzerland}

\author{P.~Grassberger}
%\email{pgrassbe@ucalgary.ca}
\affiliation{Complexity Science Group, Department of Physics and Astronomy, University of Calgary, Calgary, Alberta, Canada T2N 1N4 2}

\author{J.~S.~Andrade~Jr.}
%\email{soares@fisica.ufc.br}
\affiliation{IfB, ETH Z\"urich, CH-8093 Z\"urich, Switzerland}
\affiliation{Departamento de F\'{\i}sica, Universidade Federal do Cear\'a, 60451-970 Fortaleza, Cear\'a, Brazil}

\author{H.~J.~Herrmann}
%\email{hans@ifb.baug.ethz.ch}
\affiliation{IfB, ETH Z\"urich, CH-8093 Z\"urich, Switzerland}
\affiliation{Departamento de F\'{\i}sica, Universidade Federal do Cear\'a, 60451-970 Fortaleza, Cear\'a, Brazil}

\begin{abstract}
We study the corrections to scaling for the mass of the watershed, the bridge line, and the optimal path crack in two and three dimensions. We disclose that these models have numerically equivalent fractal dimensions and leading correction-to-scaling exponents. We conjecture all three models to possess the same fractal dimension, namely, $d_f=1.2168\pm0.0005$ in 2D and $d_f=2.487\pm0.003$ in 3D, and the same exponent of the leading correction, $\Omega=0.9\pm0.1$ and $\Omega=1.0\pm0.1$, respectively. The close relations between watersheds, optimal path cracks in the strong disorder limit, and bridge lines are further 
supported by either heuristic or exact arguments.
\end{abstract}


\maketitle 

\section{Introduction}

The watershed, defined as the line separating adjacent drainage basins (catchments), plays a fundamental role 
in water management~\cite{Vorosmarty98,Kwarteng00,Sarangi05}, landslides \cite{Dhakal04,Pradhan06,Lazzari06,Lee06}, 
and flood prevention \cite{Lee06,Burlando94,Yang07}. From observations of watersheds in nature, 
claims about their fractality have been made already long ago~\cite{Breyer92}. More recently, watersheds 
were investigated in Refs. \cite{Fehr09,Fehr11,Fehr11b} where their self-similarity was shown numerically 
for both natural and artificial landscapes. 

A fractal dimension consistent with the one of watersheds was also found for optimal path cracks in the 
limit of strong disorder. 
Optimal path cracking has been introduced by Andrade {\it et al.} \cite{Andrade09,Oliveira11,Andrade11} as 
a model for the evolution of successive optimal paths under constant failure. It describes, e.g., the 
breakdown of electrical or fluid flow through random media and has important applications also in other 
fields of science and technology, such as human transportation, fracture mechanics, or polymers in random 
environments, where finding the optimal path is a challenge 
\cite{Mezard84,Ansari85,Huse85,Huse85a,Kirkpatrick85,Kardar86,Kardar87,Perlsman92,Kertesz93,Perlsman96,Havlin05}. 

\begin{figure}[b]
\includegraphics[width=\columnwidth]{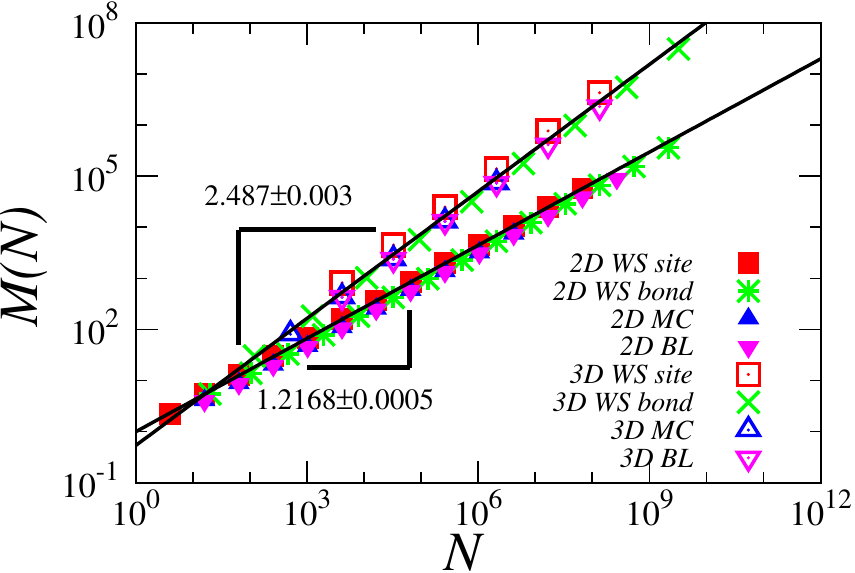}
\caption{(color online) Mass $M$ of the watershed (WS site/bond), the main crack (MC), and the bridge line (BL) as a function of the system size $N$, defined as the number of sites (bonds) in the system, in both two and three dimensions. The error bars are much smaller than the symbols. The lines show the fractal dimensions obtained in this work.
\label{massFig}
}
\end{figure}

The last of the three problems mentioned in the title, related with {\it ranked percolation} (RP) was recently introduced by Schrenk {\it et al.} \cite{Schrenk12} as a model where the creation of a spanning cluster is systematically delayed. They found that the set of ``bridge bonds" (i.e. bonds that finally lead to spanning clusters) has a fractal dimension very close to that of watersheds and of the optimal path cracks in strong disorder (see Fig.~\ref{massFig}).

The appearance of the same fractal dimension in three seemingly very different models calls on the one hand 
for a theoretical explanation, and on the other hand for more precise numerical estimates. On the theoretical
side, we might point out that the watershed (WS), the optimal path crack in strong disorder (OPC), 
and the bridge line (BL) in RP are all sets of sites or bonds that split the system into two distinct parts and 
seem conceptually related (although not identical) to classical percolation. 
Yet, despite these similarities and the broad relevance of the models, no detailed studies of the relation 
between them are available. 

Finally we should mention that also relations to other physical models have been proposed, such as optimal 
paths \cite{Andrade11,Cieplak94,Cieplak96,Porto97,Porto99}, the shortest path in loopless invasion 
percolation \cite{Cieplak96}, the infinite cluster in multiple invasion percolation \cite{Araujo04}, and 
the surface of the infinite cluster in explosive percolation \cite{Araujo10,Schrenk11}.

%%%%%%%%%%%%%%%%%%%%%%%%%%%%%%%%%%%%%%%%%%%%%%%%%%%%%%%%%%%%%%%%%%%%%%%%%%%%%%%%
%%%%%%%%%%%%%%%%%%%%%%%%%%%%%%%%%%%%%%%%%%%%%%%%%%%%%%%%%%%%%%%%%%%%%%%%%%%%%%%%
\begin{table*}
\caption{Number of samples used to obtain the average mass of the bridge line (BL), the watershed (WS sites/bonds), and the main crack (MC) for different system sizes $L$ in two- and three-dimensional systems. For the numerical analysis of corrections to scaling it is important to use high-precision data. Therefore, we focused on obtaining best possible statistics for the lattice sizes listed here, instead of increasing the number of different system sizes.\label{TABsampling}}
\begin{tabular}{l|c|c|c|c|c|c|c|c}
\hline
\hline
\multicolumn{1}{c|}{$L$} & \multicolumn{2}{|c|}{BL} & \multicolumn{2}{|c|}{WS site} & \multicolumn{2}{|c|}{WS bond} & \multicolumn{2}{|c}{MC} \\
\hline
\hspace*{1.9cm}& \hspace*{0.7cm}2D\hspace*{0.7cm} & \hspace*{0.7cm}3D\hspace*{0.7cm} & \hspace*{0.7cm}2D\hspace*{0.7cm} & \hspace*{0.7cm}3D\hspace*{0.7cm} & \hspace*{0.7cm}2D\hspace*{0.7cm} & \hspace*{0.7cm}3D\hspace*{0.7cm} & \hspace*{0.7cm}2D\hspace*{0.7cm} & \hspace*{0.7cm}3D\hspace*{0.7cm} \\
\hline
\hspace*{0.3cm}4 & $1.01\times10^{11}$ & $1.00\times10^{10}$ & $2.51\times10^{8}$ & $1.61\times10^{10}$ & $1.61\times10^{10}$ & $1.61\times10^{10}$ & $1.00\times10^{8}$ & $1.00\times10^{8}$ \\
\hspace*{0.3cm}8 & $1.20\times10^{10}$ & $1.00\times10^{9}$ & $1.37\times10^{8}$ & $2.01\times10^{9}$ & $8.05\times10^{9}$ & $2.01\times10^{9}$ & $1.00\times10^{8}$ & $1.00\times10^{8}$ \\
\hspace*{0.3cm}16 & $1.13\times10^{10}$ & $1.00\times10^{9}$ & $1.06\times10^{8}$ & $2.51\times10^{8}$ & $2.01\times10^{9}$ & $2.51\times10^{8}$ & $1.00\times10^{8}$ & $1.00\times10^{6}$ \\
\hspace*{0.3cm}32 & $3.09\times10^{9}$ & $1.34\times10^{8}$ & $7.19\times10^{7}$ & $3.14\times10^{7}$ & $5.03\times10^{8}$ & $3.14\times10^{7}$ & $5.90\times10^{7}$ & $1.00\times10^{4}$ \\
\hspace*{0.3cm}64 & $9.74\times10^{8}$ & $1.67\times10^{7}$ & $4.31\times10^{7}$ & $3.93\times10^{6}$ & $1.25\times10^{8}$ & $3.93\times10^{6}$ & $1.00\times10^{5}$ & $450$ \\
\hspace*{0.3cm}128 & $1.01\times10^{9}$ & $2.09\times10^{6}$ & $3.48\times10^{7}$ & $1.96\times10^{6}$ & $1.03\times10^{9}$ & $4.91\times10^{5}$ & $1.00\times10^{5}$ & $146$ \\
\hspace*{0.3cm}256 & $8.57\times10^{8}$ & $2.62\times10^{5}$ & $2.14\times10^{7}$ & $2.45\times10^{5}$ & $2.59\times10^{8}$ & $1.22\times10^{5}$ & $30000$ & -- \\
\hspace*{0.3cm}512 & $2.12\times10^{8}$ & $1.31\times10^{5}$ & $8.14\times10^{6}$ & $30720$ & $6.48\times10^{7}$ & $1.22\times10^{5}$ & $10400$ & -- \\
\hspace*{0.3cm}1024 & $2.68\times10^{7}$ & $4608$ & $2.31\times10^{6}$ & -- & $4.24\times10^{7}$ & $65536$ & $1310$ & -- \\
\hspace*{0.3cm}2048 & $1.95\times10^{6}$ & -- & $5.39\times10^{5}$ & -- & $1.03\times10^{7}$ & -- & $146$ & -- \\
\hspace*{0.3cm}4096 & $5.15\times10^{5}$ & -- & $1.38\times10^{5}$ & -- & $2.70\times10^{6}$ & -- & -- & -- \\
\hspace*{0.3cm}8192 & $1.28\times10^{5}$ & -- & $33573$ & -- & $7.14\times10^{5}$ & -- & -- & -- \\
\hspace*{0.3cm}16384 & $56847$ & -- & -- & -- & $1.76\times10^{5}$ & -- & -- & -- \\
\hspace*{0.3cm}32768 & -- & -- & -- & -- & $33248$ & -- & -- & -- \\
\hline
\hline
\end{tabular}
\end{table*}

In this paper we explore the relation between the main crack (MC) of the optimal path crack in strong disorder
\cite{Andrade09,Andrade11,Oliveira11} and the bridge line of RP \cite{Schrenk12}. But we shall also explore the 
relations between several definitions of watersheds \cite{Fehr09,Fehr11,Fehr11b}, since the exact definition of 
a watershed is not unique, and different definitions turn out to be closely related to different subsets of the 
other three problems. We present improved estimates 
of the fractal dimensions, made possible by studying in detail the corrections to scaling for two- and 
three-dimensional systems with uncorrelated disorder. Due to the numerical difficulty in obtaining sufficient 
statistics, we omit a discussion of the surface of the infinite cluster in discontinuous (explosive) 
percolation \cite{Araujo10,Schrenk11}. For all models, the fractal dimension $d_f$ is defined through the scaling 
of the mass $M$, corresponding to the number of sites or bonds in the object, with the linear system size $L$,
\begin{equation}\label{fractaldimGeneral}
M\sim L^{d_f}\,\, .
\end{equation}
Due to the finite system size, corrections to scaling arise \cite{Fisher70, Sur76, Huynh11} that may mask the true 
asymptotic behavior. Hence, the estimated $d_f$ can be improved by describing the size dependence of the mass as
\begin{equation}\label{massAnsatz}
M_L = L^{d_f} C_L \,\, ,
\end{equation}
where the general form for the corrections to scaling $C_L$ is
\begin{eqnarray}\label{ctsGeneral}
C_L &=& a_{00} + a_{01} L^{-1} + a_{02} L^{-2} + a_{03} L^{-3} + \dots \nonumber\\
&& + a_{11} L^{-\Omega_1} + a_{12} L^{-\Omega_1-1} + a_{13} L^{-\Omega_1-2} + \dots \nonumber\\
&& + a_{21} L^{-\Omega_2} + a_{22} L^{-\Omega_2-1} + a_{23} L^{-\Omega_2-2} + \dots \nonumber\\
&& + a_{n1} L^{-\Omega_n}+\dots \,\, ,
\end{eqnarray}
with non-universal coefficients ($a_{ij}$).
The exponents fulfill $\Omega_1 < \Omega_2 < ... < \Omega_n$ and are non-analytic (non-integer).
They are usually independent on the geometry of the lattice and only depend on the dimensionality \cite{Fisher70,Sur76}.
Finding the same non-analytic corrections-to-scaling exponents for all three models will give another hint for the close relation between them. But, in general, the precise estimation of corrections to scaling is a difficult task.
Numerical studies typically measure the leading correction exponent, a sub-leading correction exponent, or an effective exponent arising from the sum of two or more correction-to-scaling terms \cite{Caracciolo05}.
Hence, a reliable estimate of the leading correction exponent depends on both the method and the precision of the data.
Since in practice it is not reasonable to attempt a fitting with many terms of the form shown in Eq.~(\ref{ctsGeneral}), we truncate the sum of correction terms as discussed in detail below.
We first have a look at the functional form of the corrections to scaling that can be considered for the individual models given the available statistics, using a simple fitting and checking which amplitudes in Eq.~(\ref{ctsGeneral}) are small.
Using this and truncating terms with an exponent $\geq 3$, we define our effective corrections-to-scaling \emph{ansatz}.
By defining a fit quality, we identify the leading correction exponent (highest maximum of the quality) and obtain a highly accurate estimate for the fractal dimension $d_f$.
The largely improved estimate of $d_f$ is the main focus of our numerical study, rather than obtaining the corrections with precision.
We cross check the obtained results with a careful analysis of the local logarithmic slopes as suggested by Ziff \cite{Ziff99,Ziff11}.
This method uses the fact that for large enough system sizes the higher order terms are negligible, such that the local logarithmic slope of the corrections to scaling should converge to the leading correction exponent. 

The paper is organized as follows.
In Sec.~\ref{SECmodels} we describe the models.
Section \ref{SECcts} introduces the corrections to scaling and summarizes the obtained results.
The relations between the models are discussed in Sec.~\ref{SECarguments} and conclusions are drawn in Sec.~\ref{SECconclusions}.

\section{\label{SECmodels}Models}

In the following, we give a brief overview of the watershed (WS), optimal path cracking (OPC), and ranked percolation (RP), focusing on the role of percolation in the numeric procedures used to determine the watershed, the main crack (MC), and the bridge line (BL). For simplicity, the description is given for two-dimensional systems (square lattices), where they lead to lines. The extension of the discussed models to higher dimensions, where they lead to (hyper)surfaces, is straightforward and has been done in Refs.~\cite{Schrenk12,Oliveira11,Fehr11b}.

\subsection{Watersheds} \label{water}

Watersheds are the lines separating adjacent drainage basins and play a fundamental role in many 
fields \cite{Vorosmarty98,Kwarteng00,Sarangi05,Dhakal04,Pradhan06,Lazzari06,Lee06,Burlando94,Yang07}. 
Although the intuitive notion of a watershed seems obvious, the choice of a precise definition 
is rather subtle. Indeed, in the previous literature (see \cite{Vincent91} for a review) several
definitions have been used, none of which seems optimal. Moreover, as we shall see, the choice of the 
most efficient algorithm for simulating a watershed depends on the precise definition, and different definitions 
-- although corresponding to the same ``macroscopic" objects -- are more or less directly related to the 
other two problems discussed in this paper.

Following \cite{Fehr09,Fehr11,Fehr11b}, we shall discuss in the present paper two main 
definitions, the {\it bond model} and the {\it site model}, and in addition a variant of the latter, the 
{\it great wall model} (called {\it flooding method} in  \cite{Fehr09}). As we shall see, the natural 
algorithm for the bond model is one where we follow 
the run-off from top to bottom, while the natural algorithms for the site models `flood' the catchment areas 
from their outlets to the top.

We consider uncorrelated artificial landscapes mapped on a square lattice of size $L\times L$ as {\it digital elevation maps}, 
where each site $i = (x,y)$ represents a small square area. The height $h_i$ at each site $i$ is drawn randomly from 
a common distribution in such a way that $h_i > 0$. The precise form of the distribution is irrelevant, provided 
it is continuous so that, with probability one, $h_j\neq h_i, \,\, \forall_{j\neq i}$. Boundary conditions are 
periodic in the horizontal direction, but free vertically. Thus water can run across the lateral sides in both directions
(depending on which of the neighboring sites is higher), while it can only flow outwards from the top ($y=L-1$) and 
bottom ($y=0$). The latter could be modeled more explicitly by adding two more rows (with $y=L$ and $y=-1$) where all 
sites have height $h=0$, and which act as {\it sinks}. The parts of the landscape that drain to either of these 
two sinks are their {\it catchment basins}, while the line separating the two catchment basins is the watershed.

Water flows always from a higher site to a lower one, but the bond and site models correspond to different 
assumptions how this happens in detail. In the bond model, water flows from any site {\it only to its lowest
neighbor}, while it flows to {\it all neighbors} in the site model. Thus, each site belongs in the bond model
to a unique catchment area, and the watershed must be formed by {\it bonds of the dual lattice} which cut bonds 
that join sites in different 
catchment basins. It is easy to see that a watershed defined this way must be a single connected and unbranched 
path that has no loops except for the fact it is periodic in the horizontal direction (and is thus one big loop). 
Moreover, determining the catchment basin for any site is trivial: one just has to follow the unique run-off path.

In contrast, sites do not have unique run-off paths in the site model. Let us call a site with more than one 
lower neighbor a \emph{diversion site}. At each diversion site, the run-off path branches, so that the total 
run-off pattern of any site is a tree. Moreover, branches of this tree might end in both sinks, in which case 
the site cannot be in either catchment basin. Such sites must belong thus to the watershed, while sites which drain 
into one unique sink form the catchment basins. Finally, two adjacent sites $i$ and $j$ cannot be 
in different catchment basins (since either $h_i<h_j$ or $h_i>h_j$). Therefore the entire watershed must be 
formed by a single loopless and unbranched chain of sites, that is connected in the sense that adjacent sites must 
be either nearest or next-nearest neighbors.

While it is in principle possible to follow the entire run-off trees in case of the site model, it is not very 
practical and easy. Thus it is more efficient to determine the watershed by a {\it flooding} algorithm, 
where the catchment areas are determined by moving inward \& upward from the sinks. Below we shall describe 
two such algorithms that differ in details. On the other hand, for the bond model it is very efficient and 
easy to follow the run-off, as described in \cite{Fehr09}. We first determine the catchment basins for the sites
on a search line 
$(x=0,y)$ with $y=0,1,2\ldots$. The first ones will drain to the bottom. After we have found the first site 
draining to the top, we have also the first bond in the watershed. Starting from this bond we can then 
construct the entire watershed recursively, by following the run-off paths from the sites adjacent to one 
of its endpoints.

For the site model we flood {\it simultaneously} two invasion percolation clusters growing 
inward from the top and bottom rows. Let us call $B_t(h)$ and $B_b(h)$ the boundaries of these clusters, when 
the flood has height $h$. More precisely, $B_t(h)$ ($B_b(h)$) is the set of all sites $i$ with height $h_i>h$, and 
with at least one neighbor $j$ having $h_j<h$ and being in the top (bottom) cluster. Starting with $h=0$, we 
increase $h$ continuously, each time incorporating a boundary site into the corresponding cluster, as soon as 
it gets flooded -- provided this site does not belong to {\it both} boundaries. A site belonging to both 
boundaries obviously drains into both basins and is thus part of the watershed. 

When reaching the first site on the watershed, we have two options.
In one, we flood it like any other site, but take care that any site draining into it must also be in the watershed. Thus, when increasing $h$ further, we have to distinguish between sites that get flooded from neighbors that all belong to the top basin, sites that get flooded only from neighbors that all belong to the bottom basin, and sites that get flooded either from both or from a site in the watershed. The first belong to the top basin, the second to the bottom basin, and the third to the watershed.
The algorithm stops when the entire landscape is flooded.
This gives the site model proper, and is meant whenever we speak of the `site model' in the following 
sections. 

Alternatively, we can prevent sites on the watershed from being flooded by increasing their height to a value larger 
than any other $h_i$ in the entire landscape. In this way the two floods are kept separated, and we can 
continue flooding without any further modification. The algorithm stops when the entire landscape is 
flooded except for the watershed sites. These sites form then a connected wall (or dam),
whence the name {\it great wall model}. We will not present data obtained with this algorithm directly, but 
it is most closely related to the models discussed in the next two subsections.

The mass $M$ of the watershed (WS) is defined as the number of bonds (sites) forming the watershed. Notice 
that we do not consider the watershed as a three-dimensional object (with height as third dimension), but 
as 2-dimensional, see Eq.~(\ref{fractaldimGeneral}).

\subsection{Optimal Path Crack}

The optimal path crack (OPC) was introduced by Andrade {\it et al.} \cite{Andrade09,Andrade11,Oliveira11} and is obtained in the following way. We start with a square lattice of size $L$ using free boundary conditions in the vertical direction and periodic boundary conditions in the horizontal one. A random energy is assigned to each site and the energy of any path in the system is defined as the sum of the energy of its sites. In particular, the optimal path is the one among all paths connecting the top and bottom boundary of the system with the lowest total energy. Once the first optimal path is determined, the site in the optimal path having the highest energy is identified and removed. This is equivalent to imposing an infinite energy to this site. Next, the optimal path is calculated among the remaining accessible sites of the lattice, from which the highest energy site is again removed. The process continues iteratively until the system is disrupted and no further path can be found. The set of removed sites then defines the optimal path crack (OPC). The OPC is dependent on the type of disorder, but in the limit of strong disorder, it is localized in a single line, denoted as the main crack (MC), with mass $M$ given by the number of cracked sites. From this point on, we consider the OPC only in the limit of strong disorder and, for simplicity, just refer to it as main crack (MC).

In the strong disorder limit, the model is equivalent to the great wall model, with $h$ corresponding to the random 
energy and the main crack corresponding to the great wall.

\subsection{Ranked Percolation}

Ranked percolation is a new percolation model introduced by Schrenk {\it et al.} \cite{Schrenk12} 
in which the creation of a spanning cluster is suppressed. In this model bonds or sites are occupied randomly, 
except for \emph{bridges} that are bonds/sites which, when occupied, would create a spanning cluster, 
i.e. a cluster connecting top and bottom edges of the system. In the following, we focus solely on 
the case where bridges are never occupied (in the more general model of \cite{Schrenk12} they have 
a probability $p_b$ of being occupied that is smaller than the probability for other bonds/sites; in this 
notation, the present simulations correspond to $p_b=0$). While the original studies were done for 
bond percolation, we consider here site percolation. Similarly as in the bond case, we start with an 
empty square lattice of size $L\times L$, choose sites uniformly at random and occupy them. If two 
neighboring sites are occupied, they are considered to be connected and to belong to the same cluster. 
In contrast to standard site percolation, whenever the occupation of a site would lead to a spanning 
cluster, this \emph{bridge site} is blocked. The process proceeds until all sites are occupied or blocked 
and the system is disrupted into two parts. The separating \emph{bridge line} (BL) is formed by the set 
of bridge sites.

Cieplak, Maritan, and Banavar \cite{Cieplak94} have studied this line in a different context and argued 
that the occupation procedure is equivalent to the following: Randomly assign an energy to each site, 
rank order them by increasing energy, and occupy them according to their rank -- except when the site to 
be occupied is a bridge site. In that case the site is not occupied ever. Seen this way, it transpires that 
also ranked percolation is equivalent to the great wall model, except for the fact that sites are `flooded'
in different order and the algorithms suggested by the two models are very different. 

Finally, let us point out that the bond version of ranked percolation is not strictly equivalent to the 
bond model defined in subsection \ref{water}, but corresponds to a bond model on a slightly different 
lattice \cite{Schrenk12}. 

\section{\label{SECcts}Corrections to Scaling}

We perform extensive numerical simulations of the described models measuring the mass $M$ of the watershed (WS), the bridge line (BL), and the main crack (MC) for different (linear) system sizes $L$. For details about the considered system sizes and the corresponding number of samples, see Tab.~\ref{TABsampling}. The obtained masses are shown in Fig.\,\ref{massFig} as a function of the system size $N$, namely $N=L^d$ for sites and $N=d\, L^d-(2d-1)L^{d-1}$ for bonds (the second term arises due to the solid walls in the vertical direction), where $d$ is the dimensionality of the system. Although this is not visible in Fig.\,\ref{massFig}, the masses of the BL and MC are equal within the error bars. Those of WS site and WS bond are different from the masses of BL and MC. Nevertheless, we observe all of them to follow very similar scaling behaviors. The true asymptotic behavior for the mass scaling, Eq.~(\ref{fractaldimGeneral}), is masked by corrections to scaling arising due to finite system size \cite{Fisher70,Sur76}. Hence, the estimate of the fractal dimension $d_f$ can be improved by considering these corrections explicitly, see Eq.~(\ref{ctsGeneral}). In the following, we first analyze the general \emph{ansatz} to find the number of distinguishable correction exponents and if there are vanishingly small amplitudes. This results in simplified functional descriptions of the corrections to scaling in 2D and 3D, which are then studied by two different techniques in order to obtain highly accurate estimates of the exponents.

\subsection{\emph{Ansatz} for Corrections to Scaling}
%%%%%%%%%%%%%%%%%%%%%%%%%%%%%%%%%%%%%%%%%%%%%%%%%%%%%%%%%%%%%%%%%%%%%%%%%%%%%%%%
\begin{figure}
\includegraphics[width=\columnwidth]{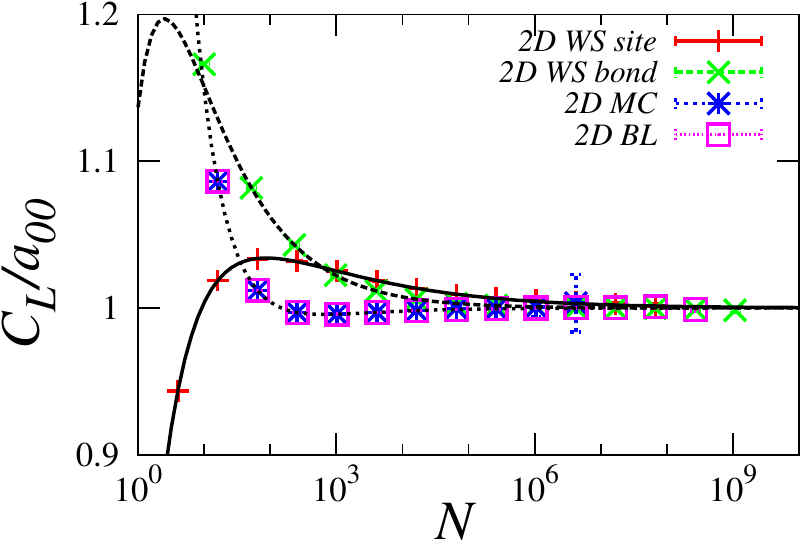}
\caption{(color online) Corrections to scaling $C_L=M/N^{d_f/d}$ of the watershed (WS site/bond), the main crack (MC), and the bridge line (BL) as a function of the system size $N$, defined as the number of sites (bonds) in the system, in two dimensions. The fractal dimension $d_f\approx1.217$, consistent with the more precise estimate obtained later, has been chosen such that $C_L$ converges to a constant value for large $N$. The error bars are typically much smaller than the symbols. The lines show fits of truncated versions of Eq.~(\ref{ctsGeneral}) to the data, which is divided here by $a_{00}$ to show the matching of the scaling behavior of the different models for large $N$.
\label{FIG_cts_intermediate}
}
\end{figure}
%%%%%%%%%%%%%%%%%%%%%%%%%%%%%%%%%%%%%%%%%%%%%%%%%%%%%%%%%%%%%%%%%%%%%%%%%%%%%%%%
To understand the structure of the data, we study least-square fits of different truncated versions of Eq.\,(\ref{ctsGeneral}) to the corrections to scaling $C_L=M/L^{d_f}$ in 2D, where $d_f\approx1.217$ has been chosen such that $C_L$ converges to a constant value for large $L$ (see Fig.~\ref{FIG_cts_intermediate}). This choice of $d_f$ is consistent with the more precise estimates obtained later. Using different numbers of exponents $\Omega_n$ and varying numbers of expansions, we find for all models that, with the current precision, we cannot resolve correction terms of an order higher than $1/L^2$. In the following, we therefore truncate the expansions by setting $a_{ij}=0\,\forall_{j>2}$. For the case of WS site, we obtain reasonable fits down to fairly small $L$ using a set of two exponents ($n=2$), yielding $\Omega_1\approx0.6$ and $\Omega_2\approx0.9$, while $a_{12}$ seems to be small and also the amplitudes of the analytic terms seem to be small and unresolvable ($a_{01}\approx0$, $a_{02}\approx0$). It is important to note that, despite these findings, $\Omega_2$ is still compatible with unity. For WS bond, MC, and BL we obtain similar results, although $a_{11}$ seems to be very small in all the three models. As shown in Fig.~\ref{FIG_cts_intermediate}, our fits match $C_L$ fairly well for the models. Hence, defining $\omega\equiv\Omega_1$ and $\Omega\equiv\Omega_2$ the (visible) corrections reduce to 
\begin{equation}\label{ctsAnsatz}
C^{2D}_L =a_{00}+a_{11}L^{-\omega}+a_{21}L^{-\Omega}+a_{22}L^{-\Omega-1} \,\, ,
\end{equation}
with $\omega\approx0.6$ and $\Omega\approx0.9$, while $a_{11}$ is large only for the WS site model. The latter fact will be discussed in section~\ref{SECarguments}.
We note that we did not find evidence of logarithmic corrections.
Figure \ref{FIG_cts_intermediate}, showing the rescaled data, confirms that the the corrections considered here capture the behavior of the data.

In 3D, we find by a similar study, that the corrections to scaling of all four models can reasonably well be described by a single correction term such that we can write
\begin{equation}\label{ctsAnsatz3D}
C^{3D}_L =a_{00}+a_{11}L^{-\Omega} \,\, ,
\end{equation}
with $\Omega\approx0.9$, but compatible with unity.
A simple least-squares fit of the \emph{ansatz} given by Eq.~(\ref{ctsAnsatz}) (Eq.~(\ref{ctsAnsatz3D}) in 3D), to the data to obtain the coefficients, $d_f$, $\Omega$, and/or $\omega$ directly can be ambiguous.
Dependent on the choice of the initial values for the fit parameters (the coefficients and exponents), a fit could even lead to an estimate of $\Omega$ or $\omega$ reflecting higher order corrections instead of the leading ones.
To overcome this and improve the accuracy, we discuss, in the following, a method that explores the parameter space by varying the exponents in a given range and analyzing the quality of the corresponding fits.
If one would attempt to fit an ansatz containing at the same time terms with variable exponents and analytic corrections to the data [formally similar to Eq.~(\ref{ctsGeneral})], interference among the terms would be possible when the variable exponent is close to unity.
The fact that $\Omega$ is close to unity, does not affect our procedure, since the corrections given in Eqs.~(\ref{ctsAnsatz}) and (\ref{ctsAnsatz3D}) do not contain analytic terms explicitly.
The results from this method are then cross checked with a second method, which allows to estimate the leading correction from the convergence of the local logarithmic slopes in the reduced mass $M\,L^{-d_f}$.
%%%%%%%%%%%%%%%%%%%%%%%%%%%%%%%%%%%%%%%%%%%%%%%%%%%%%%%%%%%%%%%%%%%%%%%%%%%%%%%%
\begin{figure}
\centering
\begin{tabular}{c}
\includegraphics[width=\columnwidth]{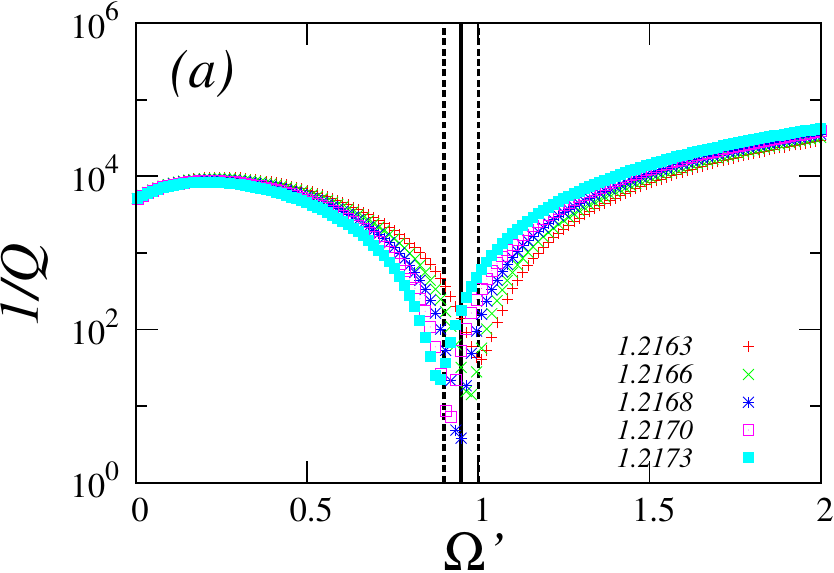}\\
\includegraphics[width=\columnwidth]{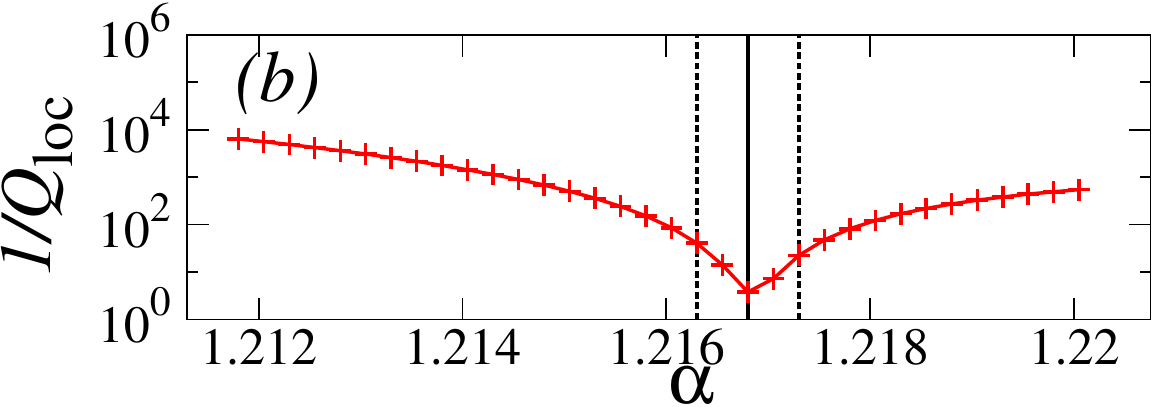}
\end{tabular}
\caption{(color online) {\bf (a)} Inverse of the quality $Q$ as a function of $\Omega'$ for different values of $\alpha$, as obtained from fits of the \emph{ansatz}, Eq.~(\ref{effAnsatz}), to the reduced mass $M\, L^{-\alpha}$ for the watershed on bonds in 2D with sizes as indicated in Tab.~\ref{TABsampling}. The vertical lines give the position of the global minimum in $1/Q$ (solid) and the estimated error (dashed). {\bf (b)} For the same system as in (a), the minimum value $1/Q_{\mathrm{loc}}$ as a function of $\alpha$ is shown, where $Q_{\mathrm{loc}}$ is obtained from curves $1/Q(\Omega')$ for a given $\alpha$, like those shown in (a). The vertical lines highlight the value of $\alpha$ at the global minimum $1/Q_{\mathrm{max}}$ (solid) and the estimate for the error (dashed). The error bars are determined from the width of the minima. The vertical lines show the estimate $d_f=1.2168\pm 0.0005$ for the fractal dimension of the watershed on bonds and the horizontal ones the corresponding leading correction $\Omega = 0.95\pm0.05$. These exponents were obtained from the analysis of a single model (WS bond). By combining the results for different models, we obtain more accurate estimates for the exponents.
\label{2DIIPbondFig}
}
\end{figure}
%%%%%%%%%%%%%%%%%%%%%%%%%%%%%%%%%%%%%%%%%%%%%%%%%%%%%%%%%%%%%%%%%%%%%%%%%%%%%%%%
\begin{figure}
\begin{center}
\includegraphics[width=\columnwidth]{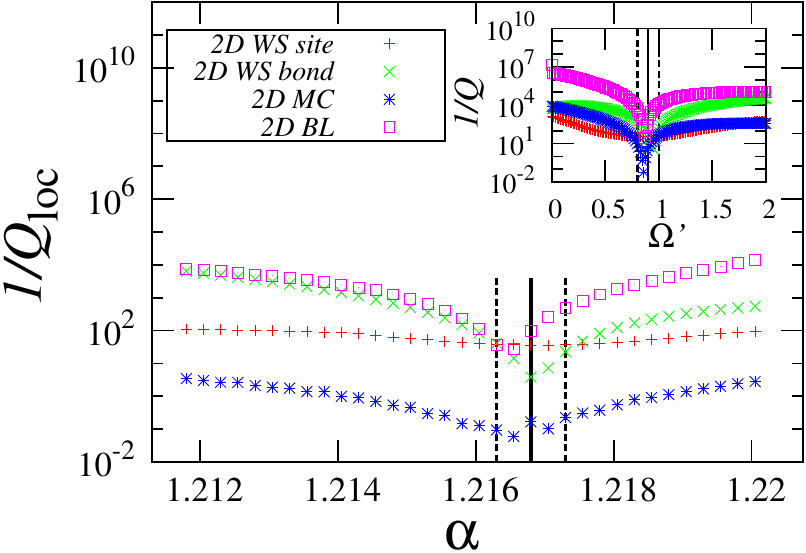}
\end{center}
\caption{(color online) Inverse of the quality at the minimum $1/Q_{\mathrm{loc}}$ as a function of $\alpha$ for the different models in 2D. The inset shows for each model the inverse of the quality $Q$ as a function of $\Omega'$ with $\alpha$ fixed to its value at the global minimum $1/Q_{\mathrm{max}}$. The vertical lines show the averages $d_f=1.2168\pm 0.0005$ and $\Omega = 0.9\pm0.1$ of the estimates for the fractal dimension and for the leading correction, respectively.
\label{2DallFig}
}
\end{figure}
%%%%%%%%%%%%%%%%%%%%%%%%%%%%%%%%%%%%%%%%%%%%%%%%%%%%%%%%%%%%%%%%%%%%%%%%%%%%%%%%

\subsection{Fit Quality Method}

The output of a fit of the \emph{ansatz} in Eq.~(\ref{ctsAnsatz}) or in Eq.~(\ref{ctsAnsatz3D}) to the data of the reduced mass $M\, L^{-d_f}$ can be sensitive to the initial conditions. We, therefore, perform a more systematic study as follows. To have a good control over the actual fitting, we use Eq.\,(\ref{massAnsatz}), (\ref{ctsAnsatz}), and (\ref{ctsAnsatz3D}) in the following form,
\begin{subequations}\label{effAnsatz}
\begin{eqnarray}
C_L(\alpha) &=& M\, L^{-\alpha} ,\\
C^{2D}_L(\alpha) &=& a_{00}+a_{11}L^{-\omega'}+a_{21}L^{-\Omega'}+a_{22}L^{-\Omega'-1} ,\,\,\,\,\,\\
C^{3D}_L(\alpha) &=& a_{00}+a_{11}L^{-\Omega'} ,
\end{eqnarray}
\end{subequations}
in 2D and 3D respectively, with fixed values of $\alpha$, $\omega'$, and $\Omega'$ and estimate the Quality $Q=n/\chi^2$, where $n$ is the number of degrees of freedom of the fit, i.e. the number of system sizes used in the data (see Tab.~\ref{TABsampling}) minus the number of fit parameters (here the number of resolvable amplitudes), and $\chi^2$ is the (weighted) mean square deviation of the fit. The quality $Q$ is a function of $\alpha$, $\omega'$, and $\Omega'$, but, as the terms of $\omega$ only have visible amplitudes for the WS site model in 2D, we drop hereafter the dependence of $Q$ on $\omega'$ and fix $\omega'=0.6$.
Since $\omega'$ is fixed, only one single correction exponent, $\Omega'$, is adjusted, avoiding fitting simultaneously multiple exponents.
Now, $Q$ should be maximal for $\alpha=d_f$ and $\Omega'=\Omega$, as the leading correction gives the dominant contribution compared to higher order ones. As a matter of convenience, we use the inverse of the quality $1/Q$, which is minimal for $\alpha=d_f$ and $\Omega'=\Omega$.
The procedure to obtain $d_f$ and $\Omega$ for a given model is to measure the inverse quality $1/Q(\alpha,\Omega')$ of a fit of the proper \emph{ansatz} to the data. We first choose a value of $\alpha$ and then derive $1/Q$ as a function of the exponent of the leading correction, scanning in the range $0<\Omega'<2$ with a step size $\delta\Omega'=0.015$. The obtained curve, see for example Fig.~\ref{2DIIPbondFig}(a) for WS bond, typically has a (local) minimum $1/Q_{\mathrm{loc}}(\alpha)$, which marks the best fit of the leading correction $\Omega_{\mathrm{loc}}(\alpha)$ for the chosen $\alpha$. The error $\Delta\Omega_{\mathrm{loc}}(\alpha)$ is estimated from the width of the minimum. In two dimensions, analyzing these minima $1/Q_{\mathrm{loc}}(\alpha)$ by varying $\alpha$ in the range $1.212< \alpha< 1.220$ with steps of size $\delta\alpha=0.00025$, yields an estimate of the global minimum $1/Q_{\mathrm{max}}$ and the fractal dimension $d_f$. The error bar in $d_f$ is also determined from the width of the minimum [see, e.g., Fig.~\ref{2DIIPbondFig}(b)]. We repeated this analysis for the watershed on sites, the main crack, and the bridge line (see Fig.~\ref{2DallFig}) and the corresponding estimates are summarized in Tab.~\ref{TABvalues}. The obtained values all agree with each other within the error bars. The ones for the MC, due to the low statistics, seem to differ more. Nevertheless, based on the similarity of the numerical values, we estimate by combining the results for all models that in 2D $d_f=1.2168\pm0.0005$ and $\Omega=0.9\pm0.1$ for all models.
The values and error bars have to be obtained by a reproducible procedure.
We used the intersection of the estimated intervals for all models (Tab.\,\ref{TABvalues}).
The value obtained for $\Omega$ is close to unity, which suggests that the leading correction (the second leading correction for WS site) is likely to be the analytic correction $\Omega=1$.

\begin{table}
\caption{The fractal dimension $d_f$ and the exponent of the leading correction $\Omega$ of the bridge line (BL) and the watershed (WS sites/bonds) for 2D and 3D, as obtained from a similar analysis as done in Fig.~\ref{2DIIPbondFig} for the WS bond case. The main crack (MC) result is only shown for 2D.
\label{TABvalues}
}
\begin{tabular}{l|c|r@{$\pm$}l|r@{$\pm$}l}
\hline
\hline
model & $\hspace*{0.525cm}d\hspace*{0.525cm}$ & \multicolumn{2}{|c|}{$d_f$} & \multicolumn{2}{|c}{$\Omega$} \\
\hline
WS bond\hspace*{0.35cm} & 2 & \hspace*{0.35cm}1.2168 & 0.0005\hspace*{0.35cm} & \hspace*{0.35cm}0.95 & 0.05\hspace*{0.35cm}\\
WS site    & 2 & 1.21705 & 0.00075 & 0.91 & 0.19 \\
BL             & 2 & 1.2166 & 0.0015 & 0.87 & 0.08 \\
MC           & 2 & 1.2166 & 0.0045 & 0.86 & 0.11 \\
WS bond & 3 & 2.4865 & 0.0025 & 0.96 & 0.10 \\
WS site    & 3 & 2.4865 & 0.0025 & 0.98 & 0.09 \\
BL             & 3 & 2.4878 & 0.0025 & 1.06 & 0.16 \\
\hline
\hline
\end{tabular}
\end{table}

\begin{figure}
\begin{center}
\includegraphics[width=\columnwidth]{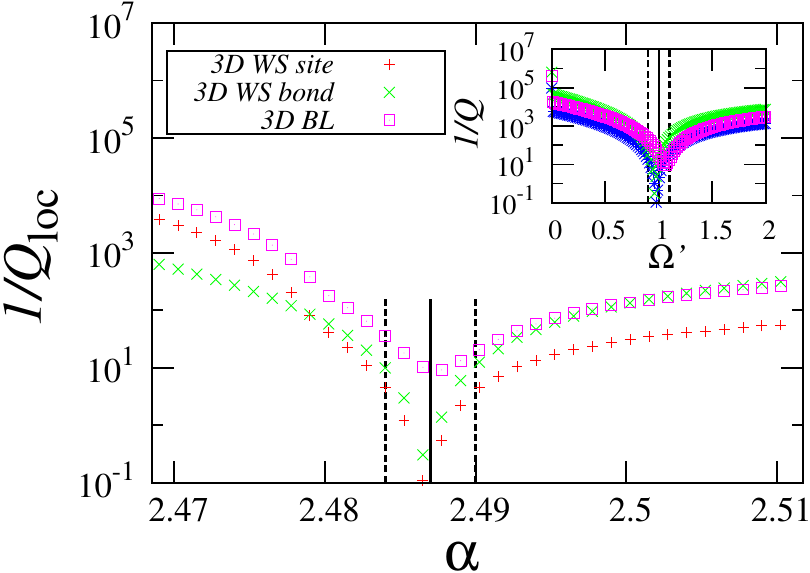}
\end{center}
\caption{(color online) Inverse of the quality at the minimum $1/Q_{\mathrm{loc}}$ as a function of $\alpha$ for the watershed on bonds, the watershed on sites, and the bridge line in 3D. The inset shows for each model the inverse of the quality $Q$ as a function of $\Omega'$ with $\alpha$ fixed to its value at the global minimum $1/Q_{\mathrm{max}}$. The vertical lines show the averages $d_f=2.487\pm 0.003$ and $\Omega = 1.0\pm0.1$ for the estimates for the fractal dimension and for the leading correction.
\label{3DallFig}
}
\end{figure}

We applied a similar analysis to the data obtained in three-dimensional systems, scanning $\Omega'$ in the range \mbox{$0<\Omega'<2$} with a step size $\delta\Omega'=0.015$ and $\alpha$ in the range \mbox{$2.450< \alpha<2.535$} with steps of size $\delta\alpha=0.0025$. As before, the detailed analysis is done like is shown in Fig.~\ref{2DIIPbondFig}). For the case of the main crack in 3D, no conclusive results could be obtained with our method, but the obtained masses are within their error bars equivalent to those measured for the bridge line. We show in Fig.~\ref{3DallFig} only the results obtained for the watershed on bonds, on sites, and the bridge line. Like in 2D, the obtained estimates for $d_f$ and $\Omega$ agree within the error bars. Therefore, we estimate $d_f=2.487\pm0.003$ and $\Omega=1.0\pm0.1$ for three dimensions. As in 2D, the value of the leading correction is likely to be the analytic one $\Omega=1$. Given this, for 2D and 3D, we also analyzed the data fixing $\Omega=1$. The obtained values for the fractal dimensions and their error bars are consistent with the ones reported in Tab.~\ref{TABvalues}, therefore, the possibility of $\Omega$ being analytical cannot be discarded.

The estimates of the fractal dimension for the different models are in agreement with the ones found in previous works for the watershed ($1.211\pm0.001$ \cite{Fehr09} and $2.48\pm0.02$ \cite{Fehr11b}), the main crack ($1.215\pm0.005$ and $2.46\pm0.05$ \cite{Andrade09,Andrade11,Oliveira11}), the bridge line ($1.215\pm0.003$ and $2.50\pm0.02$ \cite{Cieplak94,Cieplak96,Schrenk12}), and the perimeter of the infinite cluster in discontinuous percolation ($1.23\pm0.03$ \cite{Araujo10} and $2.5\pm0.2$ \cite{Schrenk11}). The value $1.211\pm0.001$ given in Ref.~\cite{Fehr09} for the fractal dimension of the watershed in two dimension seems to underestimate the error bar.

%%%%%%%%%%%%%%%%%%%%%%%%%%%%%%%%%%%%%%%%%%%%%%%%%%%%%%%%%%%%%%%%%%%%%%%%%%%%%%%%
%%%%%%%%%%%%%%%%%%%%%%%%%%%%%%%%%%%%%%%%%%%%%%%%%%%%%%%%%%%%%%%%%%%%%%%%%%%%%%%%
\subsection{Local Logarithmic Slope}

Another approach to estimate the leading correction-to-scaling exponent $\Omega$ is to calculate the local logarithmic slope of the reduced mass $C_L(\alpha)=M_{L}L^{-\alpha}$, i.e.,
\begin{equation}\label{fitAnsatz}
\Omega_{\mathrm{est}} (L,\alpha) = -\log_{2}\left( \frac{C_L(\alpha)-C_{L/2}(\alpha)}{C_{L/2}(\alpha)-C_{L/4}(\alpha)} \right) \,\, .
\end{equation}
Taking $L$ relatively large such that higher order corrections are negligible, $\Omega_{\mathrm{est}}$ converges to $\Omega$ for $\alpha=d_f$ (see, e.g., Refs.~\cite{Ziff99,Ziff11}). Due to the uncertainty $\Delta M_L$ in the average of the mass $M_L$, there are in the estimate of the local slope systematic errors of the form
\begin{eqnarray}
\Delta\Omega^2_{\mathrm{est}}(L) &=& \sum_{k=\{1,\, 2,\, 4\}}
\left(\frac{\mathrm{d}\Omega_{\mathrm{est}}}{\mathrm{d}C_{L/k}}\Delta C_{L/k}\right)^2\nonumber\\
 &=& \left( \frac{(\Delta C_L)^2+(\Delta C_{L/2})^2}{(C_{L}-C_{L/2})^2} \right)\nonumber\\
 &+& \left( \frac{(\Delta C_{L/2})^2+(\Delta C_{L/4})^2}{(C_{L/2}-C_{L/4})^2} \right)\nonumber\\
 &+& \left( \frac{(\Delta C_{L/2})^2}{(C_{L}-C_{L/2})(C_{L/2}-C_{L/4})} \right) \,\, ,\label{fitAnsatzErr}
\end{eqnarray}
where \mbox{$\Delta C_L= L^{-\alpha}\Delta M_L$}. We omitted here the $\alpha$ dependence. This error heavily depends on the precision of the single mass measurements and, therefore, statistics considerably higher than for the fit quality method are needed, especially for the larger system sizes. We focused mainly on improving the statistics for the watershed on bonds and for the bridge line, where larger systems can be addressed.
%%%%%%%%%%%%%%%%%%%%%%%%%%%%%%%%%%%%%%%%%%%%%%%%%%%%%%%%%%%%%%%%%%%%%%%%%%%%%%%%
\begin{figure}
\includegraphics[width=\columnwidth]{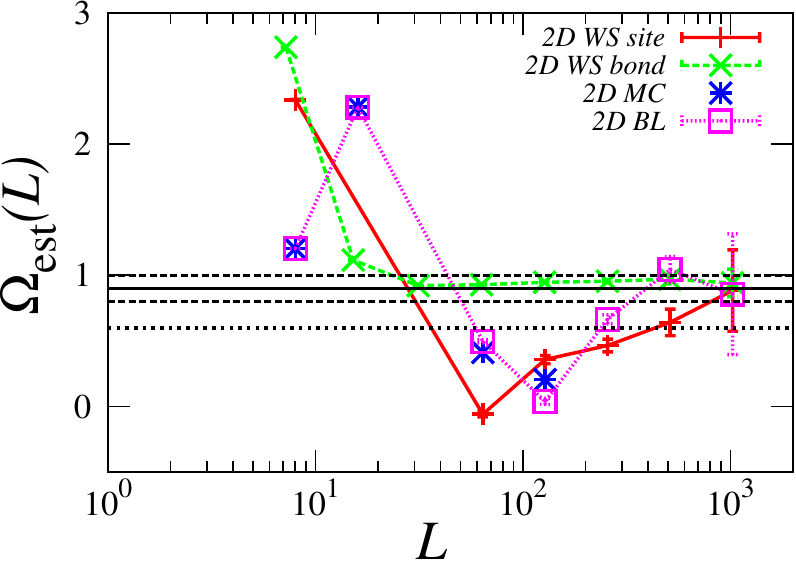}
\caption{(color online) Estimated leading correction $\Omega_{\text{est}}$ as defined in Eqs.~(\ref{fitAnsatz}) and (\ref{fitAnsatzErr}) from the mass data of the bridge line, the watershed on bonds (sites), and the main crack in 2D. The value of $\alpha$ is fixed to $1.2168$, the fractal dimension estimated by the fit quality method. For better visibility, the data of each model is shown with connecting lines and data points with $\Delta\Omega_{\mathrm{est}}>1$ have been removed. The values for the main crack (MC) are shown for comparison, but without their error bars. The horizontal lines give the value (solid) and error bar (dashed) for $\Omega$ as estimated by the fit quality method, as well as the value for $\omega$ (dotted).
\label{2DZiffFig}
}
\end{figure}
%%%%%%%%%%%%%%%%%%%%%%%%%%%%%%%%%%%%%%%%%%%%%%%%%%%%%%%%%%%%%%%%%%%%%%%%%%%%%%%%

\begin{figure} 
\includegraphics[width=\columnwidth]{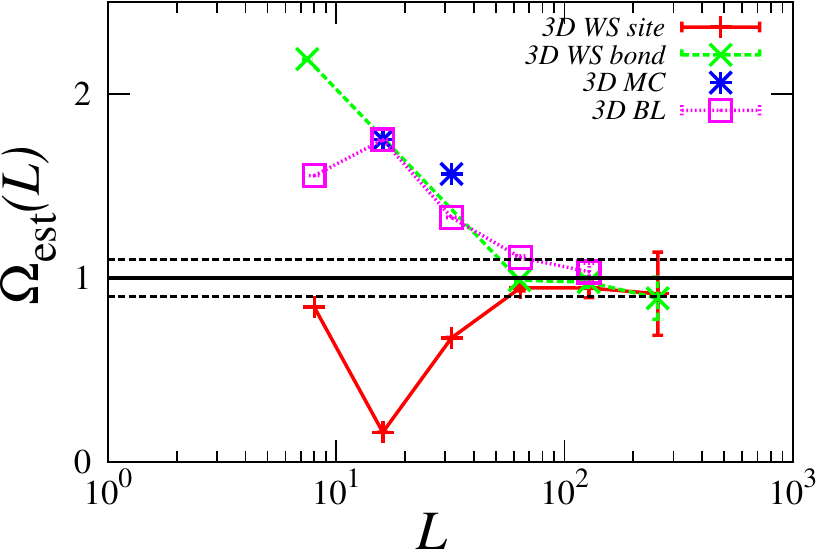}
\caption{(color online) Estimated leading correction $\Omega_{\text{est}}$ as defined in Eqs.~(\ref{fitAnsatz}) and (\ref{fitAnsatzErr}) from the mass data of the bridge line, the watershed on bonds (sites), and the main crack in 3D. The value of $\alpha$ is fixed to $2.487$, the fractal dimension estimated by the fit quality method. For better visibility, the data of each model is shown with connecting lines and data points with $\Delta\Omega_{\mathrm{est}}>1$ have been removed. The values for the main crack (MC) are shown for comparison, but without their error bars. The horizontal lines give the value (solid) and error bar (dashed) for $\Omega$ as estimated by the fit quality method.
\label{3DZiffFig}
}
\end{figure}

In Figs.~\ref{2DZiffFig} and \ref{3DZiffFig} we show $\Omega_{\mathrm{est}}$ with $\alpha=1.2168$ and $2.487$ for two- and three-dimensional systems, respectively. In both figures, only values of  $\Omega_{\mathrm{est}}$ with $\Delta\Omega_{\mathrm{est}}<1$ are shown, except those for the MC, which are shown for completeness, but without their error bars. In the limit of large $L$, we find for WS bond, MC, and BL data an agreement with the range of values for $\Omega$ obtained from the fit quality method, which corroborates our numerical results. For the WS site model in 2D $\Omega_{\mathrm{est}}$ is consistent with $\omega=0.6$, while in 3D it agrees with the other models. We cross checked also by applying other methods like, e.g., the one used in Refs.~\cite{Grassberger93,Grassberger97} and found results consistent with the ones presented here.

%%%%%%%%%%%%%%%%%%%%%%%%%%%%%%%%%%%%%%%%%%%%%%%%%%%%%%%%%%%%%%%%%%%%%%%%%%%%%%%%

\section{\label{SECarguments}Relation between the Models}

\subsection{Bridges, Cracks and Great Walls}

The numerical agreement between the bridges in ranked percolation, the optimal path cracks in the 
strong disorder limit, and the watersheds in the `great wall model' supports the claim, made in 
Sec. \ref{SECmodels}, that these models are completely equivalent. More precisely, they 
correspond to different strategies for finding the same object (the watershed, the bridge line, 
and the optimal path cracks, respectively). Since these strategies also use the random number 
generators in different ways, they lead to different statistical errors, but they give identical
scaling laws and identical corrections to scaling. 

The random occupation procedure in ranked percolation \cite{Schrenk12} can be interpreted as rank sites by increasing order in the energy and iteratively occupy them according to their position in the rank. At every step, each occupied site has a lower energy than any unoccupied one. In strong disorder, the energy of any path is dominated by the one of the site with the largest energy and, therefore, a path of occupied sites, has always lower energy than any path containing unoccupied ones. Occupying the first bridge site would lead to a spanning cluster (SC) and for the first time enable paths that connect the two opposite borders. The bridge site, as being the last occupied one in such a path, has the largest energy of all sites in it and characterizes the energy of the path. The optimal path is one of those paths, as their energy is lower than any other connecting path passing through unoccupied sites. This means that the first optimal path is cracked at the bridge site. Proceeding with the occupation of sites following the rank, the next time connecting paths are obtained is when the next bridge site is occupied. Again, the energy of the new optimal path is dominated by the energy of the current bridge site. As before, the crack appears at the bridge site. In this picture, the optimal paths always crack at bridge sites, until the system is completely disconnected. We, therefore, conjecture that the bridge line and the optimal path crack are identical.

\subsection{Interrelations between the Three Watershed Models}

As also seen from the different corrections to scaling, the relationships between the 
three watershed models are less trivial and, indeed, quite subtle.

\subsubsection{Bond and Great Wall Models}

Both in the bond model and in the great wall model, watersheds are topologically strictly 
one-dimensional closed chains. Removing even a single bond (site) from them would cut them
open, and removing two non-adjacent bonds (sites) would cut them into two disjoint pieces.
Furthermore, one can easily see that any bond in the bond watershed must be dual to a bond
adjacent to a site in the great wall, and that any such site can have at most three adjacent 
bonds corresponding to bonds in the bond watershed. This gives immediately
\begin{equation}
   M_{\rm bond} \leq 3 M_{\rm great wall}, 
\end{equation}
and therefore also the rigorous inequality $d_{f,\;\rm bond} \leq d_{f,\;\rm great wall}$.

We have no similar argument for the opposite inequality, but our numerics suggest of course 
strongly that both fractal dimensions are the same.

\subsubsection{The Site Watershed Model}

\begin{figure} 
\begin{center}
\includegraphics[width=\columnwidth]{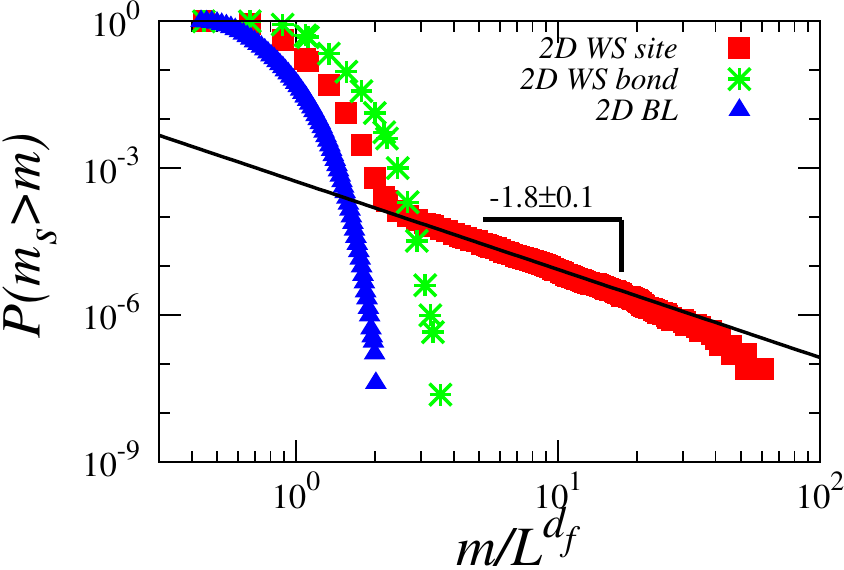}
\end{center}
\caption{(color online) Cumulative distribution $P(m_s>m)$ of the masses obtained with the WS site, WS bond, and the BL model for system size $L=128$ in 2D. The tail of the WS site case follows a power law with exponent $-1.8$.
\label{FIG_cts_massdist}
}
\end{figure}

Although one might have anticipated that the great wall model is more similar to the site 
model than to the bond model, the opposite is true. Indeed, the site model shows a strong 
anomaly that makes its finite size corrections {\it very} different, although it seems that 
it still has the same fractal dimension. This anomaly is clearly seen in Fig.~\ref{FIG_cts_massdist},
where we compare the cumulative mass distributions obtained for BL, WS bond, and WS site of 2D systems with size $L=128$.
While these distributions fall off rapidly (roughly Gaussian) for the BL and  WS bond models, we see a very 
pronounced tail in case of the WS site model.
Similar observations have been made, e.g., in Ref.~\cite{Christensen11}. 
This tail still falls off fast enough to have no effect on
the fractal dimension, but it definitely calls for an explanation.

\begin{figure} 
\includegraphics[width=0.5\columnwidth]{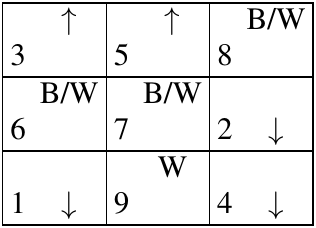} 
%\begin{center}
%\begin{tabular}{|cc|cc|cc|}
%\hline
%&$\uparrow$&&$\uparrow$&&B/W\\
%3&&5&&8&\\
%\hline
%&B/W&&B/W&&\\
%6&&7&&2&$\downarrow$\\
%\hline
%&&&W&&\\
%1&$\downarrow$&9&&4&$\downarrow$\\
%\hline
%\end{tabular}
%\end{center}
\caption{Representative system with $L=3$, where each square cell represents a site of the lattice. The numbers in the lower left corner of each cell give the heights. Letters ``B" and ``W" indicate that a site is part of the bridge line (i.e. the great wall) and of the site watershed,
respectively. Notice that the center site in the bottom row is part of the site watershed (as it is upstream of the central site), but 
is not part of the great wall, because the wall built at the center site prevents water to flow there. Arrows indicate the flow of water from 
sites belonging to the two basins.
\label{FIG_cts_differenceofmodels}
}
\end{figure}

Indeed, the watershed in the site model is not strictly one-dimensional in the topological sense, but can 
contain arbitrarily ``thick" regions where it is effectively two-dimensional. These regions correspond to 
lakes with a single outlet site, from which the water can run off towards both sinks. Their existence can 
also be deduced from the flooding algorithm used to construct the site model watershed: As explained in Sec.~\ref{SECmodels},
any site `upstream' of a watershed site has to be also on the watershed. An example of a 
very small system showing this phenomenon is given in Fig.~\ref{FIG_cts_differenceofmodels}. As 
exemplified in this figure, it follows from the algorithm that the great wall is always a subset 
of the site model watershed. Thus one has the strict inequalities
\begin{equation}
   M_{\rm site} \geq M_{\rm great wall},
\end{equation}
and $d_{f,\;\rm site} \geq d_{f,\;\rm great wall}$. Again we cannot prove rigorously the opposite inequality for the fractal dimensions, but again the numerical evidence for equality is overwhelming.

The origin of the power-law tail lies deep in the definition of the watershed on sites, namely in the fact that entire branches in the diverting runoff scheme can be part of the watershed. We will explain this here for the representative system depicted in Fig.~\ref{FIG_cts_differenceofmodels}. First, we start with the BL, occupying the sites in increasing order of the heights, so $1,\,2,\,3,\dots,9$. The first percolating cluster we would obtain when 6 is occupied, which is therefore a bridge site and the same applies to 7 and 8, while 9 just belongs to the bottom part. For WS site, we find that starting from 6 three branches develop, one to 1 and the bottom sink, another to 2, passing to 4 and reaching the bottom sink (passing 1 and directly from 4), and a third to 3 and the top sink. Hence, from 6 both sinks can be reached and it is therefore part of the watershed. The same is true for 8, two going to the top (3 and 5) and one to the bottom (2). If we now start our runoff scheme from 9, we see that initially it diverts into four branches, where three are part of the basin of the bottom sink. But the branch going upwards, is split at 7 into three sub-branches (to 2, 5, and 6), the branch from 2 again reaches the bottom sink, but the one growing from 5 is part of the top basin. Hence, 7 is part of both (or neither) basin, so it is part of the watershed and, by definition, also its parent 9 has to be considered part of the watershed. Similarly, this can be deduced from the sub-branch to 6. The watershed of this system, therefore, consists of the BL and an \emph{overhang} of one additional site. In general, such \emph{overhangs} can be larger than one site but all bridge sites are always part of the watershed.
%%%%%%%%%%%%%%%%%%%%%%%%%%%%%%%%%%%%%%%%%%%%%%%%%%%%%%%%%%%%%%%%%%%%%%%%%%%%%%%%
%%%%%%%%%%%%%%%%%%%%%%%%%%%%%%%%%%%%%%%%%%%%%%%%%%%%%%%%%%%%%%%%%%%%%%%%%%%%%%%%
%%%%%%%%%%%%%%%%%%%%%%%%%%%%%%%%%%%%%%%%%%%%%%%%%%%%%%%%%%%%%%%%%%%%%%%%%%%%%%%%

As we have conjectured, the main crack and the bridge line are identical, such that discussing the relation of WS site and BL is equivalent to discuss the relation of WS site and MC. Considering the elevations of a landscape to correspond to energies, e.g. potential energy, its watershed and its optimal path crack can be compared. We have defined that a site belongs to the watershed, when the invasion percolation clusters grown from two lower nearest neighbor sites do reach the opposite borders (catchments). As both clusters, by definition, do not cross the watershed, the watershed site separating the two, has a larger height (energy) than any site in both clusters. Therefore, for each watershed site (also the overhangs) there are always paths which consist solely of sites with lower energy than the watershed site, connecting it to either border. In the strong disorder limit, the energy of these paths is dominated by the largest local energy, i.e. the energy of the watershed site. Hence, every path connecting the two borders has at least the energy of the site where it crosses the watershed and the optimal path, the one of lowest energy, crosses the watershed at its lowest site. From the same arguments it follows that it is then the watershed site which is removed by the OPC model. After this, the next optimal path will cross the watershed at the next site in increasing order in the energy and cracks at the watershed as well. Until the final disruption of the system, in strong disorder, every crack appears at the watershed site. Hence, the MC is also a subset of the watershed.

From these findings it follows that the power-law tail in the mass distribution for WS site arises due to the existence \emph{overhangs}. We know from previous studies, that the finite-size cut-off of distributions which follow a power law can heavily affect the scaling behavior of the moments of this distribution \cite{Fehr11b}. 
The average mass $M$ is the first moment of the mass distribution $P(m_s)$ (the derivative of the distribution shown in Fig.~\ref{FIG_cts_massdist}) and, therefore, its scaling behavior is affected by the cut-off $L^2$ of its power-law tail. As we based our analysis of the corrections to scaling on $M$, also $C_L$ might be affected.
We observe the upper cut-off of the tail to scale with $L^2$ and the lower cut-off to scale with $L^{d_f}$. Therefore, the functional form of the tail of the cumulative distribution is given by $P(m_s>m)\propto m^{-1.8}L^{1.8d_f}$.
To quantify the contributions of the overhangs to $C_L$, we derive here, similar as it was done in Ref.~\cite{Fehr11b}, the contribution of the power-law tail between its cut-offs $C_\text{tail}$ which scales as
\begin{eqnarray}
C_\text{tail}&\sim& \int^{L^2}_{L^{d_f}} m_s P(m_s)\mathrm{d}m_s \nonumber  \  \  ,\\
C_\text{tail}&\sim& \int^{L^2}_{L^{d_f}} m_s \left(\left.\frac{\mathrm{d}}{\mathrm{d}m}\right|_{m=m_s}P(m_s>m)\right)\mathrm{d}m_s \nonumber  \  \  ,\\
C_\text{tail}&\sim& \int^{L^2}_{L^{d_f}} m_s \left(\left.\frac{\mathrm{d}}{\mathrm{d}m}\right|_{m=m_s}m^{-1.8}L^{1.8\,d_f}\right)\mathrm{d}m_s \nonumber  \  \  ,\\
C_\text{tail}&\sim& L^{d_f}(L^{-0.6}-\text{const})  \  \  ,
\end{eqnarray}
what leads to a contribution of order $L^{-0.6}$ to the corrections to scaling of WS site. Although it is only a rough estimate, the similarity of this contribution to the value we found for the leading correction ($\omega\approx0.6$) is striking. The other models have no such overhangs and therefore the corresponding amplitude is very small. Together with the fact that for these other models the amplitudes of the $\omega$ correction are small, this suggests that this term in $C_L$ of WS site only arises due to the overhangs. Apart from this we find the corrections to scaling of all models to be in agreement with each other. Furthermore, in 3D no such power-law tail is observed for the watershed on sites and all models hence have similar distribution of masses.

\section{\label{SECconclusions}Conclusion}

We obtained from a correction-to-scaling analysis, with high precision, an estimate for the fractal dimension of the watershed on bonds (WS bond), the watershed on sites (WS site), the bridge line (BL), and the main crack (MC). We found these fractal dimensions to be, within the error bars, in agreement with each other. All models have within error bars the same leading correction-to-scaling exponent in 2D (second leading exponent for WS site) and in 3D. These results are also corroborated by the analysis of the local logarithmic slopes in the limit of large system sizes. We estimate for all models $d_f=1.2168\pm0.0005$ and $\Omega=0.9\pm0.1$ in two dimensions and $d_f=2.487\pm0.003$ and $\Omega=1.0\pm0.1$ in three dimensions. The equivalence between the models is also supported by either heuristic or exact arguments. Furthermore, we give an explanation for the origin of the leading correction for WS site in 2D. The estimated values agree with the fractal dimensions obtained in previous studies for the watershed \cite{Fehr09,Fehr11,Fehr11b}, the optimal path crack \cite{Andrade09,Andrade11,Oliveira11}, and the bridge line \cite{Cieplak94,Cieplak96,Schrenk12}, as well as with the ones found for the perimeter of the infinite cluster in discontinuous percolation ($1.23\pm0.03$ \cite{Araujo10} and $2.5\pm0.2$ \cite{Schrenk11}). It would be interesting to know if this perimeter also obeys the same corrections to scaling as we have found.

\begin{acknowledgments}
We acknowledge useful discussions with C. Moukarzel.
We thank the ETH Risk Center for financial support.
This work has been supported by the Swiss National Science Foundation (grant number 200021-126853).
We thank CNPq, CAPES, FUNCAP, INCT-SC, and the CNPq/FUNCAP-Pronex grant for financial support.
\end{acknowledgments}

\bibliography{cts}

\end{document}